 \documentstyle[12pt,epsfig,twoside,fleqn,espcrc1]{article}

\newcommand{\AmS}{{\protect\the\textfont2
  A\kern-.1667em\lower.5ex\hbox{M}\kern-.125emS}}

\title{Probing the high-density behavior of nuclear symmetry energy with high-energy
radioactive beams} 
\author{Bao-An Li\footnote{email: Bali@astate.edu}\address{Department of Chemistry and Physics, \\
Arkansas State University, P.O. Box 419,\\
State University, AR 72467-0419, USA}}

\begin{document}

% typeset front matter
\maketitle

\begin{abstract}
Central collisions induced by high energy radioactive beams can be used as a novel 
means to obtain crucial information about the high density ({\rm HD}) behaviour of 
nuclear symmetry energy. This information is critical for understanding several key issues 
in astrophysics. Within an isospin-dependent hadronic transport 
model using phenomenological equations of state ({\rm EOS}) 
for dense neutron-rich matter, we investigate several experimental probes of 
the HD behavior of nuclear symmetry energy, such as, the $\pi^-$ to $\pi^+$ ratio, 
neutron-proton differential flow and its excitation function. 
Measurements of these observables will provide the first terrestrial 
data to constrain stringently the HD behaviour of nuclear symmetry energy and thus 
also the {\rm EOS} of dense neutron-rich matter. 
\end{abstract}

\section{INTRODUCTION}
Nuclear reactions using rare isotopes has opened up several new frontiers in 
nuclear sciences\cite{tanihata,liudo,lkb98}. In particular, high energy heavy 
rare isotopes to be available at the future Rare Isotope Accelerator (RIA) and 
the new GSI accelerator facility provide a unique opportunity to explore 
novel properties of dense neutron-rich matter 
that was not in reach in terrestrial laboratories before. This exploration 
will reveal crucial information about the ${\rm EOS}$ of neutron-rich matter.
To understand the latter and its astrophysical implications, such as,  
the origin of elements, structure of rare isotopes and properties of 
neutron stars are presently among the most important goals of nuclear 
sciences. The ${\rm EOS}$ of neutron-rich matter of isospin asymmetry 
$\delta\equiv (\rho_n-\rho_p)/(\rho_n+\rho_p)$ can be written as
\begin{equation}\label{ieos}
e(\rho,\delta)= e(\rho,0)+E_{sym}(\rho)\delta^2
\end{equation}
within the parabolic approximation (see e.g., \cite{lom}), 
where $e(\rho,0)$ is the energy per nucleon in isospin symmetric 
nuclear matter and $E_{sym}(\rho)$ is the symmetry energy at density $\rho$. 
To study the $E_{sym}(\rho)$ has been a longstanding goal of extensive research 
with various microscopic and/or phenomenological models over the last few decades. 
The predicted results on the density dependence, especially at 
high densities, are extremely diverse and often contradictory.
Theoretical results can be roughly classified into 
two groups, i.e., a group where the $E_{sym}(\rho)$ rises monotonously and 
one in which it falls with the increasing density above about twice the 
normal nuclear matter density. Above $3\rho_0$, even the sign of the symmetry energy is unclear. 
The fundamental cause of the extremely uncertain 
HD behaviour of $E_{sym}(\rho)$ is the complete lack of terrestrial laboratory 
data to constrain directly the model predictions. Recently, several promising 
observables were proposed to extract the HD behavior of the symmetry energy using
high energy radioactive beams\cite{li02}.
\begin{figure}[htp] 
%\vspace{-1.cm}
\centering \epsfig{file=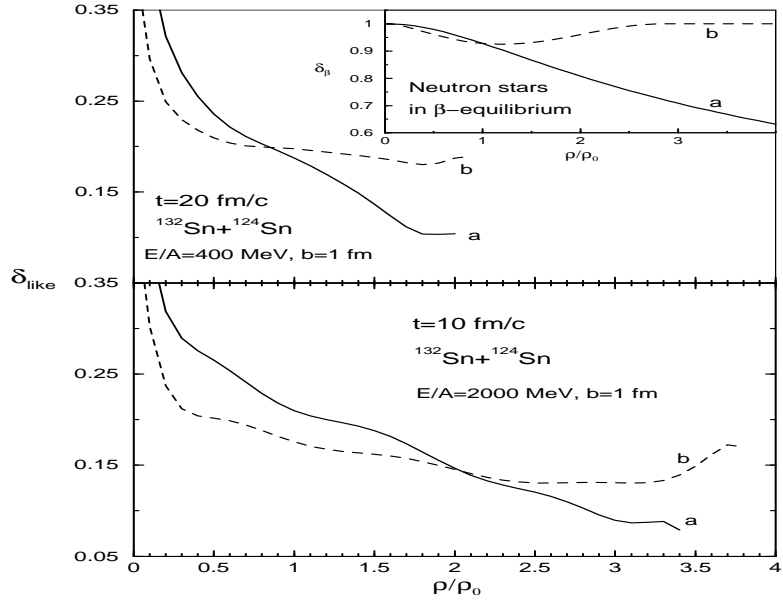,width=8cm,height=10cm,angle=-90} 
\vspace{0.1cm}
\caption{Upper window: the isospin asymmetry-density correlations at t=20 fm/c
and $E_{beam}/A=400$ MeV in the central $^{132}Sn+^{124}Sn$ reaction with the
nuclear symmetry energy $E^a_{sym}$ and $E^b_{sym}$, respectively.
Lower window: the same correlation as in the upper window but 
at 10 fm/c and $E_{beam}/A=2$ GeV/nucleon. The corresponding correlation 
in neutron stars is shown in the insert.} 
\label{fig8}
\end{figure}
\section{CORRELATION BETWEEN DENSITY AND ISOSPIN ASYMMETRY IN NEUTRON STARS AND HEAVY-ION COLLISIONS}
In a neutron star, several of its main properties are affected significantly by the correlation
between the density and the proton fraction $x_{\beta}$ at $\beta$ equilibrium. 
The latter is determined by\cite{lat91}    
\begin{equation}\label{fraction}
\hbar c(3\pi^2\rho x_{\beta})^{1/3}=4E_{\rm sym}(\rho)(1-2x_{\beta}).
\end{equation}   
The equilibrium proton fraction is therefore entirely determined by 
the $E_{sym}(\rho)$. We use the following two representatives 
of the symmetry energy 
\begin{equation}\label{esym}
E^a_{sym}(\rho)\equiv E_{sym}(\rho_0)u ~~~{\rm and}~~~
E^b_{sym}(\rho)\equiv E_{sym}(\rho_0)u\cdot\frac{3-u}{2},
\end{equation}
where $u\equiv\rho/\rho_0$. 
The values of $\delta_{\beta}=1-2x_{\beta}$ corresponding to 
these two forms of symmetry energy are shown in the insert of Fig.\ 1.
With the $E^{b}_{sym}(\rho)$, 
the $\delta_{\beta}$ is $1$ for $\rho/\rho_0\geq 3$, 
indicating that the neutron star has become a pure neutron matter at 
these high densities. On the contrary, with the $E^{a}_{sym}(\rho)$, 
the neutron star becomes more proton-rich as the density increases.
To see the connection between the physics governing properties of both neutron stars and 
heavy-ion collisions, we show in Fig. 1 the correlation between density $\rho$
and the isospin asymmetry $\delta_{like}$ (effective isospin asymmetry taking into account different 
charge states of $\Delta(1232)$ resonances) over the entire reaction volume at the time 
of about the maximum compression in the central $^{132}Sn+^{124}Sn$ reactions 
with E/A=400 (upper window) and 2000 (bottom window) MeV/nucleon, respectively. 
These results were obtained using an isospin-dependent hadronic transport model\cite{ibuu1}.
The overall rise of $\delta$ at low 
densities is mainly due to the neutron skins of the colliding nuclei and the 
distillated neutrons. Effects due to the different symmetry energies are more clearly 
revealed especially at high densities.
An astonishing similarity is seen in the resultant 
$\delta-\rho$ correlations for the neutron star and the heavy-ion collision. 
In both cases, the symmetry energy $E^b_{sym}(\rho)$ makes the HD nuclear 
matter more neutron-rich than the $E^a_{sym}(\rho)$ and the effect 
grows with the increasing density. Of course, this is no surprise 
since the same underlying nuclear ${\rm EOS}$ is at work in both cases. 
The decreasing $E^b_{sym}(\rho)$ above $\frac{1}{2}u_c=1.5\rho_0$ makes 
it more energetically favorable to have the denser region more neutron-rich in both neutron 
stars and heavy-ion collisions. 

\section{$\pi^-/\pi^+$ PROBE OF THE $E_{sym}(\rho)$}
It is well known that the $\pi^-/\pi^+$ ratio in heavy-ion collisions depends strongly on the isospin asymmetry of the
reaction system. It is also qualitatively easy to understand why this dependence
can be used to extract crucial information about the {\rm EOS} of neutron-rich matter. On one hand, within the $\Delta$
resonance model for pion production from first-chance independent nucleon-nucleon collisions\cite{stock}, the primordial $\pi^-/\pi^+$ ratio 
is $(5N^2+NZ)/(5Z^2+NZ)\approx (N/Z)^2$. It is thus a direct measure of the isospin 
asymmetry $(N/Z)_{dense}$ of the dense matter in the participant region of heavy-ion collisions. It was shown that the $(N/Z)_{dense}$ 
is uniquely determined by the high density behaviour of the nuclear symmetry energy\cite{li02}. Therefore, the $\pi^-/\pi^+$ ratio
can be used to probe sensitively the {\rm EOS} of neutron-rich matter. 
On the other hand, within the statistical model for pion production\cite{nature}, the $\pi^-/\pi^+$ ratio is proportional to 
${\rm exp}\left[(\mu_n-\mu_p)/T\right]$, 
where T is the temperature, $\mu_n$ and $\mu_p$ are the chemical potentials of neutrons and protons, respectively. 
At modestly high temperatures ($T\geq 4$ MeV), the difference in the neutron and proton chemical potentials can be 
written as\cite{li02a}
\begin{equation}
\mu_n-\mu_p=V^n_{asy}-V^p_{asy}-V_{Coulomb}+T\left[{\rm ln}\frac{\rho_n}{\rho_p}+\sum_m\frac{m+1}{m}b_m(\frac{\lambda_T^3}{2})^m(\rho^m_n-\rho^m_p)\right],
\end{equation}
where $V_{Coulomb}$ is the Coulomb potential for protons, $\lambda_T$ is the thermal wavelength of a nucleon and $b'_m$s are the inversion 
coefficients of the Fermi distribution function. The difference in neutron and proton symmetry 
potentials $V^n_{asy}-V^p_{asy}=2v_{asy}(\rho)\delta$, where the function 
$v_{asy}(\rho)$ is completely determined by the density-dependence of the symmetry energy\cite{li02}. It is seen that the kinetic 
part of the difference $\mu_n-\mu_p$ relates directly to the isospin asymmetry $\rho_n/\rho_p$ or $\rho_n-\rho_p$.
Thus within the statistical model too, the $\pi^-/\pi^+$ ratio is sensitive to the $(N/Z)_{dense}$. Moreover, the value of $\pi^-/\pi^+$ ratio 
is affected by the competition of the symmetry and Coulomb potentials which all depend on the isospin asymmetry of the reaction system.
At relatively low temperatures there is a good chance to extract crucial information about the
symmetry potential.

\begin{figure}[htp] 
\vspace{-0.2cm}
\centering \epsfig{file=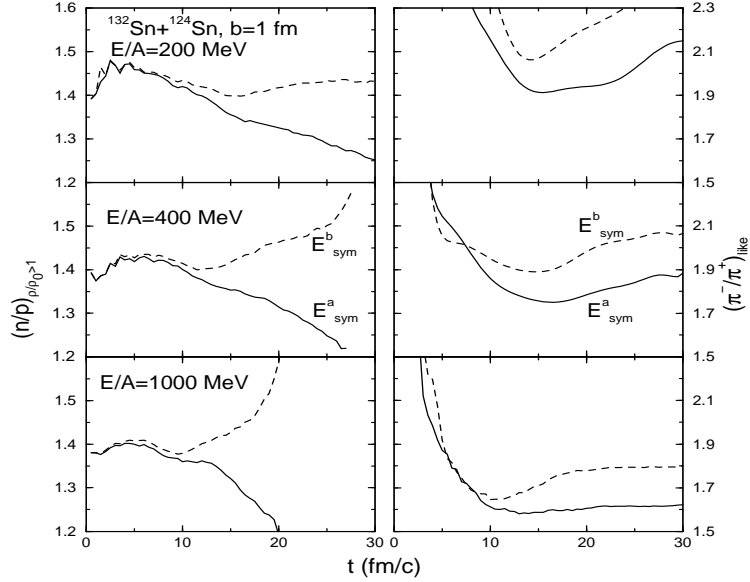,width=8cm,height=10cm,angle=-90} 
\vspace{0.1cm}
\caption{The evolution of neutron/proton ratio of dense region (left) 
and $\pi^-/\pi^+$ ratio (right) in the central $^{132}Sn+^{124}Sn$ reactions 
at $E_{beam}/A=200, 400$ and 1000 MeV.} 
\label{fig2}
\end{figure}

Shown in Fig.\ 2 are the isospin asymmetry of the high density region 
$(n/p)_{\rho\geq \rho_0}$ (left) and the $(\pi^-/\pi^+)_{like}$ ratio (right)
\begin{equation}
(\pi^-/\pi^+)_{like}\equiv \frac{\pi^-+\Delta^-+\frac{1}{3}\Delta^0}
{\pi^++\Delta^{++}+\frac{1}{3}\Delta^+}
\end{equation} 
as a function of time for the central $^{132}Sn+^{124}Sn$ reaction at 
beam energies from 200 to 1000 MeV/nucleon. This ratio naturally 
becomes the final $\pi^-/\pi^+$ ratio at the freeze-out 
when the reaction time $t$ is much longer 
than the lifetime of the delta resonance $\tau_{\Delta}$.
The $(\pi^-/\pi^+)_{like}$ ratio is rather high in the early 
stage of the reaction because of the large numbers of neutron-neutron 
scatterings near the surfaces where the neutron skins of the colliding nuclei overlap. 
It is seen that a variation of about 
30\% in the $(n/p)_{\rho/\geq \rho_0}$ due to the different $E_{sym}(\rho)$ 
results in about 15\% change in the final $\pi^-/\pi^+$ ratio.
It has thus an appreciable response factor of 
about 0.5 to the variation of HD n/p ratio and is approximately 
beam energy independent. Therfore, the $(\pi^-/\pi^+)_{like}$ ratio 
is a sensitive probe of the HD behaviour of nuclear symmetry energy.

\begin{figure}[htp] 
\vspace{-1.3cm}
\centering \epsfig{file=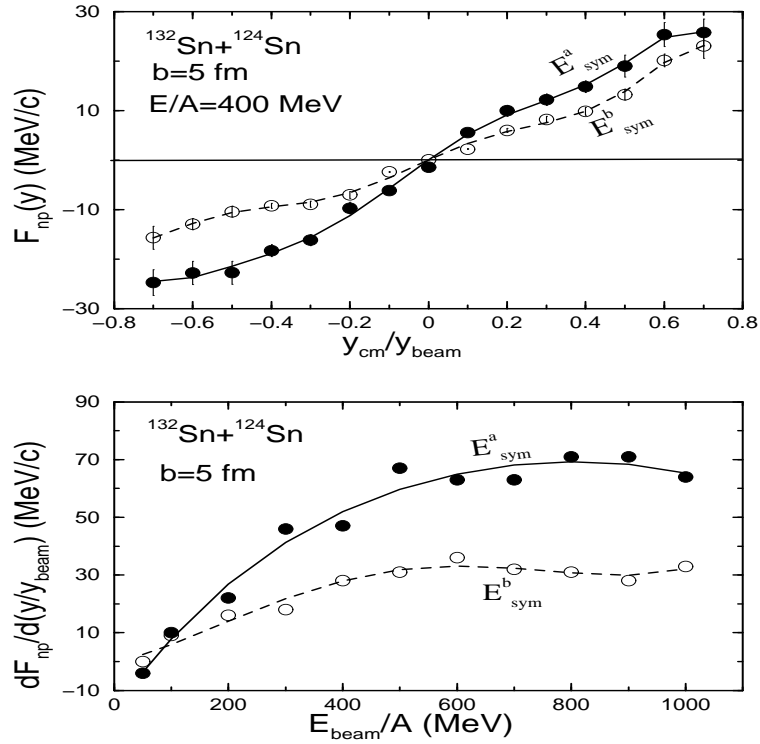,width=10cm,height=10cm,angle=-90} 
\vspace{0.1cm}
\caption{The neutron-proton differential collective flow
in the mid-central $^{132}Sn+^{124}Sn$ reactions at $E_{beam}/A=400$ MeV (upper window) 
and the excitation function of flow parameter (lower window) with the nuclear symmetry energy 
$E^a_{sym}$ and $E^b_{sym}$, respectively.} 
\label{fig15}
\end{figure}

\section{NEUTRON-PROTON DIFFERENTIAL FLOW PROBE OF THE $E_{sym}(\rho)$}
The neutron-proton differential collective flow is measured
by\cite{li00} 
\begin{equation}
F_{np}(y)\equiv\frac{1}{N(y)}\sum_{i=1}^{N(y)}p_{x_i}\tau_i,
\end{equation}
where $N(y)$ is the total number of free nucleons at the rapidity $y$, 
$p_{x_i}$ is the transverse momentum of particle $i$ in the reaction 
plane, and $\tau_i$ is $+1$ and $-1$ for neutrons and protons, respectively.
The free nucleons are identified as those having local nucleon densities less
than $1/8\rho_0$. The $F_{np}(y)$ combines constructively the in-plane transverse 
momenta generated by the isovector potentials while reducing 
significantly influences of the isoscalar potentials of both neutrons 
and protons. Thus, it can reveal more directly the HD behaviour 
of $E_{sym}(\rho)$ in high energy heavy-ion collisions. A typical 
analysis for mid-central $^{132}Sn+^{124}Sn$ reactions at 400 MeV/nucleon 
is made in the upper window of Fig.\ 3. A clear 
signature of the HD behaviour of $E_{sym}(\rho)$ 
appears at both forward and backward rapidities. The slope 
$dF_{np}/d(y_{cm}/y_{beam})$ at mid-rapidity is different by about 
a factor of 2 in reactions at $E_{beam}\geq 200$ MeV/nucleon as shown in the lower
window of Fig.\ 3. This large effect can be easily observed by using available 
detectors at several heavy-ion facilities in the world. 

\section{SUMMARY}
In summary, the HD behaviour of nuclear symmetry energy has been 
puzzling physicists for decades. In this work, central collisions induced by 
high energy radioactive beams are proposed as a novel means 
to solve this longstanding problem. Within an isospin dependent hadronic 
transport model using two representative density dependent symmetry energy 
functions predicted by many body theories, several experimental 
probes for the HD symmetry energy are studied. Among them, the $\pi^-/\pi^+$ ratio, 
neutron-proton differential collective flow and its excitation function are found 
most promising. Future comparisons between the experimental data and model calculations 
will constrain stringently the HD behaviour of nuclear symmetry energy. 

This work was supported in part by the National Science Foundation under grant No. PHY-0088934.

\end{document}